\documentstyle[prl,aps,multicol]{revtex}
\begin{document}
\title{Turbulent Dynamos and Magnetic Helicity}
\draft
\input{epsf}
\author{Hantao Ji}
\address{Princeton Plasma Physics Laboratory, Princeton University,
	P.O. Box 451, Princeton, NJ 08543}
\date{Received 30 April 1999; Published in Physical Review Letters 
Vol. 83, 3198 (1999)}

\maketitle

\begin{abstract} 

It is shown that the turbulent dynamo $\alpha$-effect converts magnetic 
helicity from the turbulent field to the mean field when the turbulence is 
electromagnetic while the magnetic helicity of the mean-field is transported 
across space when the turbulence is electrostatic or due to the 
electron diamagnetic effect. In all cases, however, the dynamo effect 
strictly conserves the total helicity except for resistive effects and 
a small battery effect. Implications for astrophysical situations, 
especially for the solar dynamo, are discussed.

\end{abstract}


\begin{multicols}{2}
\dimen100=\columnwidth
\setlength{\columnwidth}{3.375in}
\parskip=0in

Magnetic fields are observed to exist not only in the planets and the 
stars~\cite{Parker79} but essentially everywhere in the universe, 
such as the interstellar medium in galaxies and even in clusters of 
galaxies~\cite{Zweibel}.
The origin of these cosmical magnetic fields has been explained mainly 
by dynamo theory~\cite{Moffatt}, which is one of the most active 
research areas across 
multiple subdisciplines of physics. In particular, generation of an electromotive 
force (EMF) along a mean field by turbulence, or the well-known $\alpha$
effect~\cite{Parker55}, is an essential process in amplifying 
large-scale magnetic fields~\cite{Krause}. Experimentally, the $\alpha$ 
effect has been observed in toroidal laboratory plasmas~\cite{Ji96}.

Recently, there has been growing awareness that a topological constraint 
on the observed magnetic field, the conservation of magnetic helicity, 
may play an important role in solar flare evolution~\cite{Rust}.
This follows the success of Taylor in explaining the observed magnetic
structures in laboratory plasmas by conjecturing the same constraint
during relaxation~\cite{Taylor86}.
Magnetic helicity, a measure of the \lq\lq knottedness\rq\rq 
and the \lq\lq twistedness" of magnetic fields~\cite{Woltjer,Berger}, is 
closely related to the dynamo effect. Indeed, the $\alpha$ effect drives 
parallel current which twists up the field lines, thus increasing magnetic 
helicity on large scales. As a matter of fact, almost all the observed
large scale cosmical poloidal (or meridional) magnetic fields, either in 
their dipolar or quadrupolar forms, have linkage with strong toroidal
(or azimuthal) fields, leading to finite magnetic helicity.

One simple yet important question arises: how exactly is magnetic helicity 
affected by the dynamo process? Can magnetic helicity of the large-scale field 
be created by the dynamo process or merely be transported across space? Motivated 
by Taylor's conjecture, early studies~\cite{Boozer86} showed that the $\alpha$ 
effect only transports helicity of the large-scale field across space without affecting 
the total helicity, as supported by laboratory measurements\cite{Ji95a}.
However, a contradicting conclusion was drawn in a recent 
study~\cite{Seehafer}, which showed that the $\alpha$
effect locally converts helicity from the turbulent field to the mean field, 
as supported by statistical and numerical studies on inverse 
helicity cascading to large scales~\cite{Pouquet,Stribling}.
Answers to the questions raised by this contradiction are in demand since they 
would reveal the nature of the dynamo effects and clarify the effectiveness or limitations 
of the magnetic helicity concept in determining the evolution of solar and laboratory 
plasmas in which the the dynamo plays a role.

In this Letter, it is shown that both conclusions, {\it i.e.} creation or transport of the 
large-scale magnetic helicity by the $\alpha$ effect, are valid depending 
on the nature of the turbulence which drives the dynamo effect. When the turbulence is 
electromagnetic, the $\alpha$ effect converts helicity from the 
turbulent, small-scale field to the mean, large-scale field. On the other hand, when 
the turbulence is electrostatic or due to the electron diamagnetic 
effect, the $\alpha$ effect transports the mean-field helicity across space without 
dissipation. In all cases, however, the $\alpha$ effect strictly conserves 
the total helicity except for resistive effects and a small battery effect.
Implications for
astrophysical situations, especially for the solar dynamo, are discussed.

In order to include other possible dynamo effects in a plasma, we revisit the mean-field
electrodynamics~\cite{Krause}
using the generalized Ohm's law (ignoring the electron inertial term)~\cite{Spitzer}
\begin{equation}
{\bbox{E}} + {\bbox{v}} \times {\bbox{B}}
-{\bbox{j}} \times {\bbox{B}}/ en
+\mbox{\boldmath $\nabla$} P_e / en = \eta {\bbox{j}},
\label{GeneralizedOhm}
\end{equation}
where $n$ is the electron density and $P_e$ the electron pressure.
Every quantity $x$ is divided into a mean part $\overline x \equiv \langle x\rangle $,
averaged over ensembles or space, and a turbulent part $\widetilde x$: $x =
\overline x + \widetilde x$.
Then the mean and turbulent versions of the Ohm's law become
\begin{eqnarray}
\overline {\bbox{E}} + \overline {\bbox{v}}_e \times \overline 
{\bbox{B}} +{\mbox{\boldmath $\nabla$} \overline P_e / en} 
+\mbox{\boldmath $\mathcal{E}$} & = & \eta \overline {\bbox{j}}
\label{Ohm_mean}\\
\widetilde {\bbox{E}} + \widetilde {\bbox{v}}_e \times \overline {\bbox{B}}
+ (\overline {\bbox{v}}_e + \widetilde {\bbox{v}}_e) \times \widetilde {\bbox{B}}
- \mbox{\boldmath $\mathcal{E}$}
+{\mbox{\boldmath $\nabla$} \widetilde P_e / en} 
& = & \eta \widetilde {\bbox{j}},
\label{Ohm_turb}
\end{eqnarray}
where ${\bbox{v}}_i$ (${\bbox{v}}_e$) is the ion (electron) flow velocity
and the relations ${\bbox{v}} \approx {\bbox{v}}_i$ and 
${\bbox{j}}=en({\bbox{v}}_i-{\bbox{v}}_e)$ have been used.
The mean EMF $\mbox{\boldmath $\mathcal{E}$}$ is given by 
\begin{equation}
\mbox{\boldmath $\mathcal{E}$}
= \langle \widetilde {\bbox{v}} \times \widetilde {\bbox{B}} \rangle -
\langle \widetilde {\bbox{j}} \times \widetilde {\bbox{B}} \rangle /e n
\approx \langle \widetilde {\bbox{v}}_e \times \widetilde {\bbox{B}} \rangle .
\label{EMF}
\end{equation}
(Small battery-like effects such as $\langle \widetilde n \mbox{\boldmath $\nabla$} 
\widetilde P_e\rangle /e\overline n^2$ are neglected; see discussions later.)
The appearance of ${\bbox{v}}_e$ only on the RHS of Eq.(\ref{EMF}) is
consistent with Ohm's law being a force balance on {\it electrons}.

The parallel component of $\mbox{\boldmath $\mathcal{E}$}$, or the
$\alpha$-effect~\cite{Parker55}, along the mean field is of interest via Eq.(\ref{Ohm_turb}):
\begin{eqnarray}
\mbox{\boldmath $\mathcal{E}$} \cdot \overline {\bbox{B}}
& = & \langle \widetilde {\bbox{v}}_e \times \widetilde {\bbox{B}} \rangle  \cdot \overline {\bbox{B}}
= - \langle (\widetilde {\bbox{v}}_e \times \overline {\bbox{B}} ) \cdot \widetilde {\bbox{B}}\rangle 
\nonumber \\
& = & \langle \widetilde {\bbox{E}} \cdot \widetilde {\bbox{B}}\rangle 
+\langle \mbox{\boldmath $\nabla$} \widetilde P_e \cdot \widetilde {\bbox{B}}\rangle /e \overline n
-\eta \langle \widetilde {\bbox{j}} \cdot \widetilde {\bbox{B}}\rangle ,
\label{EMF_c}
\end{eqnarray}
where the last term diminishes in the limit of small resistivity~\cite{Pouquet,Diamond} and shall
be discussed later.
The first term $\langle \widetilde {\bbox{E}} \cdot \widetilde {\bbox{B}}\rangle $ 
represents the contribution to $\widetilde {\bbox{v}}_e$
from the turbulent $\widetilde {\bbox{E}} \times \overline{\bbox{B}}$
drift which is a single fluid (MHD) effect~\cite{Ji94},
while the second term, $\langle \mbox{\boldmath $\nabla$} \widetilde P_e \cdot \widetilde 
{\bbox{B}}\rangle /e n$, is the contribution from the turbulent 
electron diamagnetic drift $\mbox{\boldmath $\nabla$} \widetilde P_e \times \overline 
{\bbox{B}}$ which is an {\it electron} fluid effect~\cite{Ji95b}.

In general, the electric field can be split further into a curl-free part and a 
divergence-free part, often called \lq\lq electrostatic\rq\rq and \lq\lq 
electromagnetic\rq\rq, respectively: ${\bbox{E}}=-
\mbox{\boldmath $\nabla$} \phi- \partial {\bbox{A}} / \partial t$ where
${\bbox{A}}$ is the vector potential and
$\phi$ is the electrostatic potential. Then Eq.(\ref{EMF_c}) becomes
\begin{eqnarray}
\mbox{\boldmath $\mathcal{E}$} \cdot \overline {\bbox{B}}
= & - & \langle \mbox{\boldmath $\nabla$} \widetilde \phi 	
	\cdot \widetilde {\bbox{B}}\rangle 
- \langle  (\partial \widetilde {\bbox{A}} / \partial t) 
	\cdot \widetilde {\bbox{B}}\rangle  \nonumber \\
& + & \langle \mbox{\boldmath $\nabla$} \widetilde P_e \cdot \widetilde {\bbox{B}}\rangle  / e n
-\eta \langle \widetilde{\bbox{j}} \cdot \widetilde {\bbox{B}}\rangle ,
\label{EMF_d}
\end{eqnarray}
where the first three terms correspond to effects due to electrostatic,
electromagnetic, and electron diamagnetic turbulence, respectively~\cite{electrostatic}.
We shall see below that the type of turbulence is crucial in 
assessing effects of dynamo action on the magnetic helicity.

Magnetic helicity~\cite{Woltjer} in a volume $V$ is defined~\cite{gauge} by 
$ K = \int {\bbox{A}} \cdot {\bbox{B}} dV $ and its rate of change is given by
\begin{equation}
{d K \over dt } = - 2 \int {\bbox{E}} \cdot {\bbox{B}} dV 
- \int ( 2 \phi {\bbox{B}} + {\bbox{A}} \times {\partial {\bbox{A}} \over \partial t}) 
\cdot d{\bbox{S}},
\label{balance_K}
\end{equation}
where $V$ is enclosed by the surface ${\bbox{S}}$.
The integral under the volume integration represents the {\it volume} 
rate of change of 
helicity, while the integral under the surface integration represents {\it flux} of 
helicity. We note that only the volume term can
possibly create or destroy helicity, and the surface terms merely 
transport helicity across space without affecting the total helicity.
The mean helicity $\langle K\rangle $ is the sum 
of the helicity in the mean field, $K_{\rm m} = \int \overline {\bbox{A}} \cdot 
\overline {\bbox{B}} dV$, and the helicity in the turbulent field, $K_{\rm t} = \int 
\langle \widetilde {\bbox{A}} \cdot \widetilde {\bbox{B}}\rangle  dV$. From Eq.(\ref{balance_K}), 
we have
\begin{eqnarray}
{d K_{\rm m} \over dt } = 
& - & 2 \int \overline {\bbox{E}} \cdot \overline {\bbox{B}} dV 
- \int ( 2 \bar \phi \overline {\bbox{B}} +
\overline {\bbox{A}} \times {\partial \overline {\bbox{A}} \over \partial t} 
) \cdot d{\bbox{S}} \nonumber \\
{d K_{\rm t} \over dt } = 
& - & 2 \int \langle \widetilde {\bbox{E}} \cdot \widetilde {\bbox{B}}\rangle  dV 
- \int \langle 2 \widetilde \phi \widetilde {\bbox{B}}
+ \widetilde {\bbox{A}} \times {\partial \widetilde {\bbox{A}} \over
\partial t}\rangle \cdot d{\bbox{S}}, 
\nonumber
\end{eqnarray}
where substitution of $\overline{\bbox{E}}$ and $\widetilde {\bbox{E}}$ by 
Eqs.(\ref{Ohm_mean}) and (\ref{Ohm_turb}) yields
\begin{eqnarray}
\overline {\bbox{E}} \cdot \overline {\bbox{B}}
&=& \eta \overline {\bbox{j}} \cdot \overline {\bbox{B}}
- \mbox{\boldmath $\mathcal{E}$} \cdot \overline {\bbox{B}}
- \mbox{\boldmath $\nabla$} \cdot \left(\overline P_e \overline {\bbox{B}} / en
\right) \label{mean_K} \\
\langle \widetilde {\bbox{E}} \cdot \widetilde {\bbox{B}}\rangle 
&=& \eta \langle \widetilde {\bbox{j}} \cdot \widetilde {\bbox{B}}\rangle 
+ \mbox{\boldmath $\mathcal{E}$} \cdot \overline {\bbox{B}}
- \mbox{\boldmath $\nabla$} \cdot \left(\langle \widetilde P_e \widetilde {\bbox{B}}\rangle / en \right).
\label{turb_K}
\end{eqnarray}
It might be concluded that the dynamo effects
convert helicity from the turbulent field to the mean field since
$\mbox{\boldmath $\mathcal{E}$} \cdot \overline {\bbox{B}}$ appears on both equations
but with opposite signs~\cite{Seehafer}.
However, substitution of $\mbox{\boldmath $\mathcal{E}$} \cdot \overline {\bbox{B}}$
by Eq.(\ref{EMF_d}) in Eqs.(\ref{mean_K}) and (\ref{turb_K}), using
$\int \langle \mbox{\boldmath $\nabla$} \widetilde \phi \cdot \widetilde {\bbox{B}}\rangle  dV =
\int \langle \widetilde \phi \widetilde {\bbox{B}}\rangle  \cdot d{\bbox{S}}$ etc., yields
\begin{eqnarray}
{d K_{\rm m} \over dt } &=&
- 2 \int (\eta \overline {\bbox{j}} \cdot \overline {\bbox{B}}
+ \eta \langle \widetilde {\bbox{j}} \cdot \widetilde {\bbox{B}}\rangle 
+ \underbrace{\langle  {\partial \widetilde {\bbox{A}} \over \partial t}
     \cdot \widetilde {\bbox{B}}\rangle }_A) dV \nonumber \\
& & - \int (\underbrace{2 \bar \phi \overline {\bbox{B}}}_B
        -\underbrace{2{\overline P_e \overline {\bbox{B}} \over en}}_C
        +\underbrace{\overline {\bbox{A}} \times {\partial \overline {\bbox{A}} 
	\over \partial t}}_D + \underbrace{2\langle \widetilde \phi \widetilde {\bbox{B}}\rangle }_E
\nonumber \\
& & 	-\underbrace{2{\langle \widetilde P_e \widetilde {\bbox{B}} \rangle \over en} }_F) \cdot 
d{\bbox{S}} \label{mean} \\
{d K_{\rm t} \over dt } &=& 
2 \int \underbrace{\langle  {\partial \widetilde {\bbox{A}} \over \partial t} 
\cdot \widetilde {\bbox{B}}\rangle }_A dV 
- \int \underbrace{\langle \widetilde {\bbox{A}} \times {\partial \widetilde {\bbox{A}} \over
        \partial t}\rangle }_G \cdot d{\bbox{S}},
\label{turb}
\end{eqnarray}
where, in Eq.(\ref{turb}), the turbulence-induced helicity flux, such as
$\langle \widetilde \phi \widetilde {\bbox{B}} \rangle$, have been cancelled
by the corresponding terms in $\langle \widetilde {\bbox{E}} \cdot \widetilde
{\bbox{B}} \rangle$. (In fact, Eq.(\ref{turb}) can be derived more simply
without involving $\widetilde \phi$ or $\widetilde P_e$ terms.)

A brief discussion is useful here for each term of these equations.
The term $D$ is responsible for the most common source of helicity for
a toroidal laboratory plasma, in which a transformer supplies poloidal (toroidal) flux 
to be linked with existing toroidal (poloidal) flux.
The term $B$ is responsible for the technique often called 
\lq\lq electrostatic helicity injection\rq\rq~\cite{Taylor89},
in which a voltage is applied between two ends of a flux tube. 
The same amount of helicity with the opposite sign is also injected into the space outside the system, 
which is often a vacuum region~\cite{Boozer93}.
The term $C$ has never been used to inject or change the helicity in a system.
The term $G$ represents transport of helicity in the turbulent field
by the propagation of electromagnetic waves possessing finite helicity.
One example is circularly polarized Alfv\'{e}n waves in a magnetized plasma.
In the ideal MHD limit, these waves propagate with no decay and no effects on
the mean field helicity since term $A$ vanishes. Finite dissipation or wave-particle interactions 
can result in a finite term $A$ which converts helicity from the turbulent field
to the mean field~\cite{Alfven} or vice versa~\cite{Yoshida}.

The role of the turbulent dynamo in helicity evolution depends 
critically on the nature of the turbulence.
When the turbulence is electromagnetic, i.e.
$\widetilde {\bbox{v}}_e$ is driven by an inductive electric field,
the dynamo effect generates the same amount of helicity in both
the mean and turbulent fields but with opposite signs, as seen from
term $A$. When the turbulence is electrostatic or electron 
diamagnetic, i.e. $\widetilde {\bbox{v}}_e$ is driven by an electrostatic field
or an electron pressure gradient, the dynamo action does not affect
the turbulent helicity but merely transports the mean-field 
helicity across space, as seen from the terms $E$ and $F$.
Note that in order for terms $E$ and $F$ to have a net effect on the mean-field helicity, 
the electrons must be non-adiabatical, {\it i.e.} $e \widetilde \phi/T_e \ne \widetilde n /n$,
a condition often satisfied in the laboratory.

\begin{figure} 
{\epsfxsize=3in
\epsffile{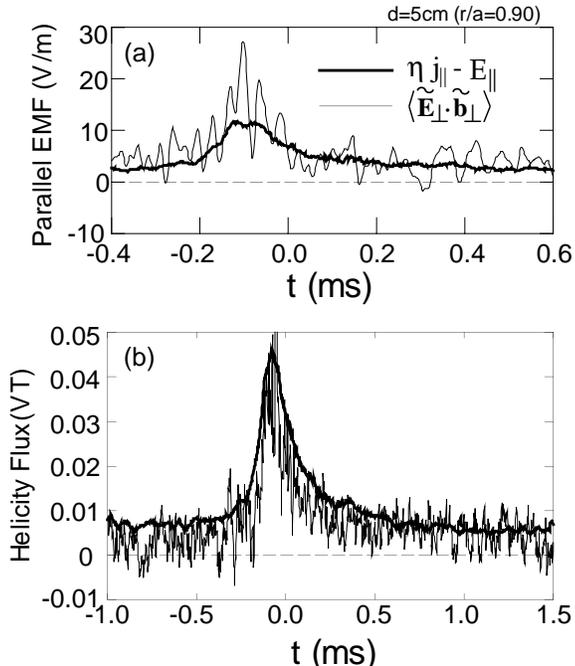}}
\caption{Measured (a) parallel EMF ($\alpha$-effect) due to electrostatic
turbulence, $\langle \widetilde {\bf E}_\perp \cdot \widetilde {\bf b}_\perp\rangle $ 
(thin line) where $\widetilde {\bf E}_\perp=-{\mbox{\boldmath 
$\nabla$}}_\perp \widetilde \phi$ and ${\bf b}={\bf B}/B$,
and (b) helicity flux (thin line) $\langle \widetilde \phi \widetilde {\bf B}\rangle $ 
in a laboratory plasma (Ref. 6). The thick lines in both (a) and (b) are the predictions
from the rest of the terms in Ohm's law and the helicity balance equation. 
The good agreements indicate that the electrostatic turbulence alone is responsible for 
both dynamo action and helicity transport.
(The $t=0$ refers to the timing of magnetic relaxation events, during which 
both the $\alpha$-effect and helicity transport are enhanced over a 
constantly working turbulent dynamo effect.)}
\end{figure}

Despite the long history of the dynamo problem, there are no generally accepted 
theories on the nature of the turbulence. It also 
has not been investigated numerically. Experimentally, however, it has been 
measured that
the turbulence responsible for the observed $\alpha$-effect in laboratory Reversed-Field Pinch (RFP)
plasmas is predominantly electrostatic~\cite{Ji94} or electron diamagnetic~\cite{Ji95b}.
In either case, the dynamo effect causes helicity transport in the mean field without 
effects on the turbulent field, consistent with theories~\cite{Boozer86} and 
experiments~\cite{Ji95a}. Figure 1 shows an example of 
measured helicity flux induced by the electrostatic turbulence 
together with the measured $\alpha$-effect in an RFP plasma.
Both measurements (thin lines) agree well with the predictions (thick lines) 
from the Ohm's law and the helicity balance equation,
indicating that the electrostatic turbulence alone is responsible for 
both dynamo action and helicity transport.

In the case of astrophysical dynamos, however, there is no 
observational evidence on the nature of the responsible turbulence.
Such knowledge would have great implications on the role of dynamo 
action in helicity evolution. A good example under
debate is the solar dynamo problem and its relationship
with the observed twisted field lines (hence the helicity) on the solar
surface~\cite{Rust94,Pevtsov} and even in the solar wind~\cite{Rust}.
It has been found that there is a preference in the sign of the observed helicity 
in each hemisphere. A generally accepted argument is that 
this helicity preference originates from the convection 
zone or even a thin layer at the bottom of the convection zone where 
the solar dynamo is believed to be operational~\cite{Gilman}. If the
turbulence is electromagnetic, magnetic helicity in the large-scale 
field will be generated while leaving the same amount of helicity with the 
opposite sign in the small-scale turbulence.
On the other hand, if the turbulence is electrostatic or electron diamagnetic,
the dynamo action will not affect helicity in the small-scale field 
but will transport or separate the large-scale helicity of one sign to one 
hemisphere while leaving the opposite helicity in the other hemisphere. 
After rising to the solar surface via buoyancy,
these large-scale structures and its associated helicity
are constantly removed from the sun by flaring.
Both mechanisms can replace the lost helicity continuously.
However, the former mechanism  conserves magnetic helicity 
locally in each hemisphere while both hemispheres need to be included
for the latter mechanism to conserve helicity.

Despite the lack of theoretical insight, we point out a general tendency in which 
the ratio of kinetic energy to magnetic energy, or the plasma beta $\beta$ in a genaral sense,
may play an important role in determining the nature of the turbulence.
When $\beta \lesssim 1$, the turbulence is prone 
to be electrostatic or electron diamagnetic, consistent with laboratory measurements.
Each field line can have a different electrostatic potential $\phi$ or electron pressure 
$P_e$ insulated by the strong magnetic field, leading to notable gradients in the 
perpendicular direction.  On the other hand, 
when $\beta \gg 1$, the turbulence becomes 
less electrostatic or electron diamagnetic due to diminishing magnetic insulation 
in the perpendicular direction and becomes more electromagnetic since the 
field lines tend to be pushed around by a much larger plasma pressure.
This conjecture is supported by a general tendency of \lq\lq reduction 
of dimensionality\rq\rq~\cite{Zank}, in which isotropic 3D turbulence reduces 
to anisotropic, 2D turbulence when a strong large-scale magnetic field 
is introduced. 

In contrast to the low-beta plasmas in the laboratory, astrophysical plasmas 
with an active dynamo usually have a beta much larger than unity. In addition to the
solar dynamo, similar situations exist for cases of the
geodynamo~\cite{Roberts} and the galactic dynamo~\cite{Kulsrud,Zweibel}. 
The aforementioned conjecture would predict a local conversion process of magnetic 
helicity by dynamo action from the turbulent field to the mean field.

Regardless of the nature of the turbulence, the total helicity 
is always conserved besides the resistive effects as per Eqs.(\ref{mean}) and (\ref{turb}). 
This can be shown more
rigorously by substituting the generalized Ohm's law Eq.(\ref{GeneralizedOhm}) into
the first term on the RHS of Eq.(\ref{balance_K}) to yield
\begin{equation}
\int {\bbox{E}} \cdot {\bbox{B}} dV = \int \eta {\bbox{j}} \cdot {\bbox{B}} dV
+ \int {\mbox{\boldmath $\nabla$} P_e \cdot {\bbox{B}} \over en} dV.
\nonumber
\end{equation}
The first term on the RHS is a resistive effect, 
which vanishes with zero resistivity.
The second term can be rewritten as
$\int (\mbox{\boldmath $\nabla$} P_e \cdot {\bbox{B}} / en) dV
= \int (T_e/e) {\bbox{B}} \cdot  d{\bbox{S}} + 
\int (T_e \mbox{\boldmath $\nabla$} n \cdot {\bbox{B}}/en) dV
= \int (T_e/e)(1+\ln n) {\bbox{B}} \cdot  d{\bbox{S}}
- \int (\ln n \mbox{\boldmath $\nabla$} T_e \cdot {\bbox{B}}/e) dV$
for which both finite gradients in density and electron
temperature (of course also in electron pressure) along the field line
are necessary conditions to change the total helicity.
However, we note that such parallel gradients, especially 
$\nabla_\parallel T_e$, are very small owing to fast electron 
flow along the field lines (with a few exceptions such as in 
laser-produced plasmas~\cite{Stamper}). Such effects, often called the battery 
effect~\cite{Parker79}, provide only a seed for magnetic 
field to grow in a dynamo process and, of course, it can be accompanied 
by small but finite magnetic helicity. 
The approximate conservation of the total helicity
during dynamo action is consistent with laboratory
observation~\cite{Ji95a}. 

Finally, it is worth commenting on a classical case of statistically stationary and homogeneous
turbulence~\cite{Krause}. In this special case, by definition, all statistical quantities of the
turbulence do not vary in time and space, leading to vanishing $dK_t/dt$ and all turbulence-induced
helicity flux: terms $E$, $F$, and $G$ in Eqs.(\ref{mean}) and (\ref{turb}). 
It follows that from Eq.(\ref{turb}), term $A$ vanishes and thus only the last 
term in Eq.(\ref{EMF_d}) survives~\cite{Seehafer94}: $\mbox{\boldmath $\mathcal{E}$} \cdot 
\overline {\bbox{B}}= -\eta \langle \widetilde{\bbox{j}} \cdot \widetilde {\bbox{B}}\rangle $.
As a result, the $\alpha$-effect, appearing as a resistive term, generates the same amount of helicity but with opposite signs
in $K_m$ and $K_t$ ~\cite{Seehafer}, but the helicity generation in 
$K_t$ is canceled out exactly by the resistive decay due to the turbulence, 
assuring $dK_t/dt=0$. 

In summary, it has been shown that the effect of turbulent dynamos
on magnetic helicity depends critically on the nature of the turbulence.
When the turbulence is electromagnetic, the $\alpha$ effect converts helicity from the 
turbulent, small-scale field to the mean, large-scale field. On the other hand, when 
the turbulence is electrostatic or due to the electron diamagnetic 
effect, the $\alpha$ effect transports the mean-field helicity across space without 
dissipation. Both mechanisms can explain the observed helicity 
preference of large-scale magnetic structures on the solar surface
but they conserve helicity in different ways.
Based on laboratory observations of turbulent dynamos,
it is conjectured heuristically that plasma beta plays an important 
role in determining the nature of the turbulence; {\it i.e.} turbulent flow is driven by
(curl-free) electrostatic electric fields or electron pressure gradient when $\beta \lesssim 1$
and by (divergence-free) electromagnetic electric fields when $\beta \gg 1$.
In all cases, however, dynamo processes conserve total helicity except for 
resistive effects and a small battery effect,
consistent with the laboratory observations.
Detailed understanding of dynamo turbulence and its effects on magnetic
helicity await further investigations not only by theories and numerical simulations but also 
by observations in space and well-controlled laboratory experiments. 

The author is grateful to Dr. R. Kulsrud, Dr. P. Diamond, and Dr. M. Yamada
for their comments, and to Dr. S. Prager and his group for RFP data.

\end{multicols}

\begin{references}
\bibitem{electrostatic} Obviously, \lq\lq pure electrostatic\rq\rq turbulence
where $\widetilde {\bbox{B}}=0$ does not have dynamo effects. 
\lq\lq Electrostatic\rq\rq is used here to refer to the situation in which
$-\langle \mbox{\boldmath $\nabla$} \widetilde \phi \cdot \widetilde {\bbox{B}}\rangle $ 
dominates over all other terms in Eq.(6).
\bibitem{gauge} When $V$ is not singly connected or the magnetic 
field is not tangent to its surface, the definition of magnetic 
helicity needs to be modified (such as the introduction of relative helicity 
in Ref. 10) for gauge invariance. However, since the 
main conclusions do not depend on these variations, the simplest definition 
is used here.
\end{references}
\end{document}